\newif\ifnature
\naturefalse

\newif\ifnabs
\nabstrue

\newcommand{\theabstract}{
\ifnabs
	In 1861, Maxwell derived two of his equations of electromagnetism by modelling a magnetic line of force
	as a `molecular vortex' in a fluid-like medium.
	Later, in 1980, Berry and colleagues conducted experiments on a `phase vortex', a wave geometry in a fluid 
	which is analogous to a magnetic line of force and also exhibits
	behaviour corresponding to the quantisation of magnetic flux. 
	Here we unify these approaches by writing down  
	a solution to the equations of motion for a compressible fluid which behaves 
	in the same way as a magnetic line of force.
	We then revisit Maxwell's historical inspiration, namely Faraday's 1846 model of light as disturbances in lines of force. 
	Using our unified model, we show that such disturbances resemble photons: they are polarised, 
	absorbed discretely,
	obey Maxwell's full equations of electromagnetism to first order, 
	and quantitatively reproduce
	the correlation that is observed in the Bell tests.
\else
	After Euler modelled light as waves in a compressible fluid,  
	Maxwell modelled a magnetic line of force as a `molecular vortex' in a fluid-like medium.
	It was later found that magnetic flux is quantised, and Berry explored a `phase vortex', a wave geometry in a fluid that is quantised in the same way. 
	Here we show these three models are related, and unify them. We write down 
	a phase-vortex solution to the equations for Euler's fluid which behaves as Maxwell described for a magnetic line of force.
	Disturbances in the lines obey Maxwell's equations of electromagnetism to first order; like photons, they are absorbed discretely and
	are consistent with the Bell tests.
\fi
}

\newcommand{\thetitle}{Maxwell's fluid model of magnetism}

\newcommand{\pbar}{\overline{\textbf{p}}}

\newif\ifonecolumnabstract
\onecolumnabstracttrue

\ifnature
	\documentclass[12pt]{article}
\else
	\documentclass[12pt,twocolumn]{article}
\fi
\pdfoutput=1
\usepackage[margin=20mm]{geometry}

\usepackage{url}
\usepackage{graphicx}
\usepackage{amssymb,amsmath}
\ifnature
	\usepackage[super,nosort]{cite} 
	\usepackage{lineno}
	\linenumbers
\else
	\usepackage{cite} 
\fi

\begin{document}

\title{\thetitle}

	\author{Robert Brady and Ross Anderson\\
	        \small University of Cambridge Computer Laboratory\\
	        \small JJ Thomson Avenue, Cambridge CB3 0FD, United Kingdom
\ifnature
\else	        
	        \\
	        \small \texttt{\{robert.brady,ross.anderson\}@cl.cam.ac.uk}
\fi
}

	\date{\today}

\date{\today}

\ifnature
\else
	\ifonecolumnabstract
		\twocolumn[
		  \begin{@twocolumnfalse}
		    \maketitle
			\begin{abstract}
		  		\theabstract
			\end{abstract}
		  \end{@twocolumnfalse}
		  ]
	\else
		\maketitle
		\section*{Abstract}
		\theabstract
	
		\section{Introduction}
	\fi
\fi

\begin{figure}[htb]
	\centering
		 \includegraphics[width=.5\textwidth]{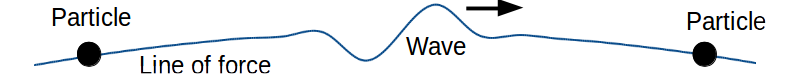}
		  \includegraphics[width=0.125\textwidth]{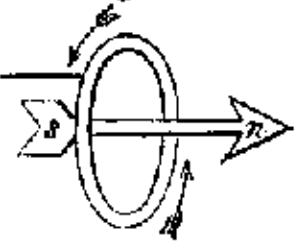}
	\caption{\em Faraday's 1846 model of light as waves in lines of force, and Maxwell's 1861 figure showing his extension to a magnetic line of force.}
	\label{fig:faraday-line-force}
\end{figure}

In 1746 Euler modelled light as waves in a frictionless compressible fluid; a century later in 1846, Faraday modelled it as vibrations in `lines of force' as in figure
\ref{fig:faraday-line-force}~\cite{euler1746opuscula,hakfoort1995optics,faraday1846rayvibrations,cantor1981conceptions}.

Fifteen years later Maxwell combined these approaches,
proposing that a magnetic line of force is a `molecular
vortex' (see the diagram from his 1861 paper in figure \ref{fig:faraday-line-force}~\cite{maxwell1855faraday,maxwell1861linesofforce,siegel2003innovation}). 
A fluid-like medium flows around the line, and centrifugal forces reduce the pressure near the centre, 
giving a 
`tension' along the axis which accounts for the forces 
between the poles of magnets.

%

Maxwell then derived two of his equations of electromagnetism. Suppose the mean momentum per unit volume of fluid is $\pbar(\textbf{x})$.
In modern notation with unit charge, the magnetic field is $\textbf{B}=\nabla \times \pbar$,
and it obeys Gauss's law for magnetism  
$\nabla.\textbf{B}=0$, since $\nabla.(\nabla \times \pbar)$ is identically zero.
Defining the magnetic flux by 
$\phi = \int \textbf{B}.d\textbf{s}$ where $d\textbf{s}$ is a surface element,
Stokes's theorem shows that $\phi=\oint \pbar.d\boldsymbol{\ell} \neq 0$
where the path
$d\boldsymbol{\ell}$ encircles the centre. 
If the fluid exerts a mean force density \textbf{E} on an external system then it must lose momentum, $\textbf{E}=-\partial \pbar/\partial t$. 
Faraday's law of induction follows immediately: $\nabla \times \textbf{E}=-\partial \textbf{B}/\partial t$.
Feynman later rediscovered a similar derivation~\cite{dyson1990feynman}. 
On later interpretations, the momentum density $\pbar$ corresponds to the magnetic vector potential.

Maxwell's magnetic line of force can be almost
any axis with mass flow around it ($\oint \pbar.d\boldsymbol{\ell}\neq 0$).
An ordinary vortex in a fluid is not a good example, since it is 
pinned to the fluid and is therefore not symmetric under Lorentz transformation~\cite{faber1995fluid}.
A better example is suggested by a series of experiments, 
starting a hundred years later, on quantised magnetic flux. 

\section{Phase vortex}

In 1961 Deaver and Fairbank, and independently Doll and N\"abauer, 
showed that magnetic flux is
quantised~\cite{deaver1961quantisedflux,doll1961experimental}.
A superconductor has a macroscopic `order parameter', written $R e^{i S}$ 
where $R$ is the amplitude and $S$ is the phase~\cite{josephson1962coupled}.
When a superconducting ring encloses $n$ quanta of magnetic flux,
the phase $S$ advances by $2 n \pi$ around it,
\begin{equation}
	\oint{\nabla S . d\boldsymbol\ell} ~=~~ 2 n \pi
\label{eq:define-phase-vortex}
\end{equation}
where $d \boldsymbol{\ell}$ is the circumference and $n$ is an integer. 

\begin{figure}[htb]
	\centering
		\includegraphics[width=0.4\columnwidth]{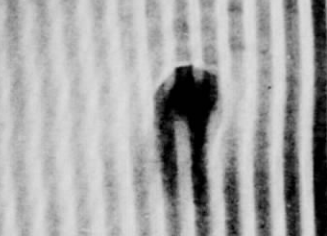}
	\caption{\em Water waves near a steady vortex. Note there is one more wave-crest above the centre
	than below it. (Courtesy Michael Berry~\cite{berry1980wavefront})}
	\label{fig:berry_1980_one_vortex}
\end{figure}

In fluid mechanics, the wave geometry in \eqref{eq:define-phase-vortex} is 
called a `phase vortex'.
In  1980 Berry, Chambers, Large, Upstill and Walmsley made a steady 
`bathtub' vortex by draining water from a tank, and sent water waves past
it as shown in figure~\ref{fig:berry_1980_one_vortex}~\cite{berry1980wavefront}. 
The waves propagate slower when they travel against the flow, producing more wavecrests above the 
centre than below it. The number of extra wavecrests depends on the vortex strength.
If the waves are continuous (apart from near the axis, where they vanish)  
the increase in phase is quantised as in (\ref{eq:define-phase-vortex}).

As well as exhibiting an analogue of flux quantisation, 
the experiment showed good agreement with the Aharonov-Bohm effect which deflects
a charged particle near a magnetic field.
Later, Fleury, Sounas, Sieck, Haberman and Al\`u made a phase vortex in sound waves 
and observed an analogue of the magnetic
Zeeman effect~\cite{fleury2014sound}.

\begin{figure}[htb]
	\centering
		\includegraphics[width=0.4\columnwidth]{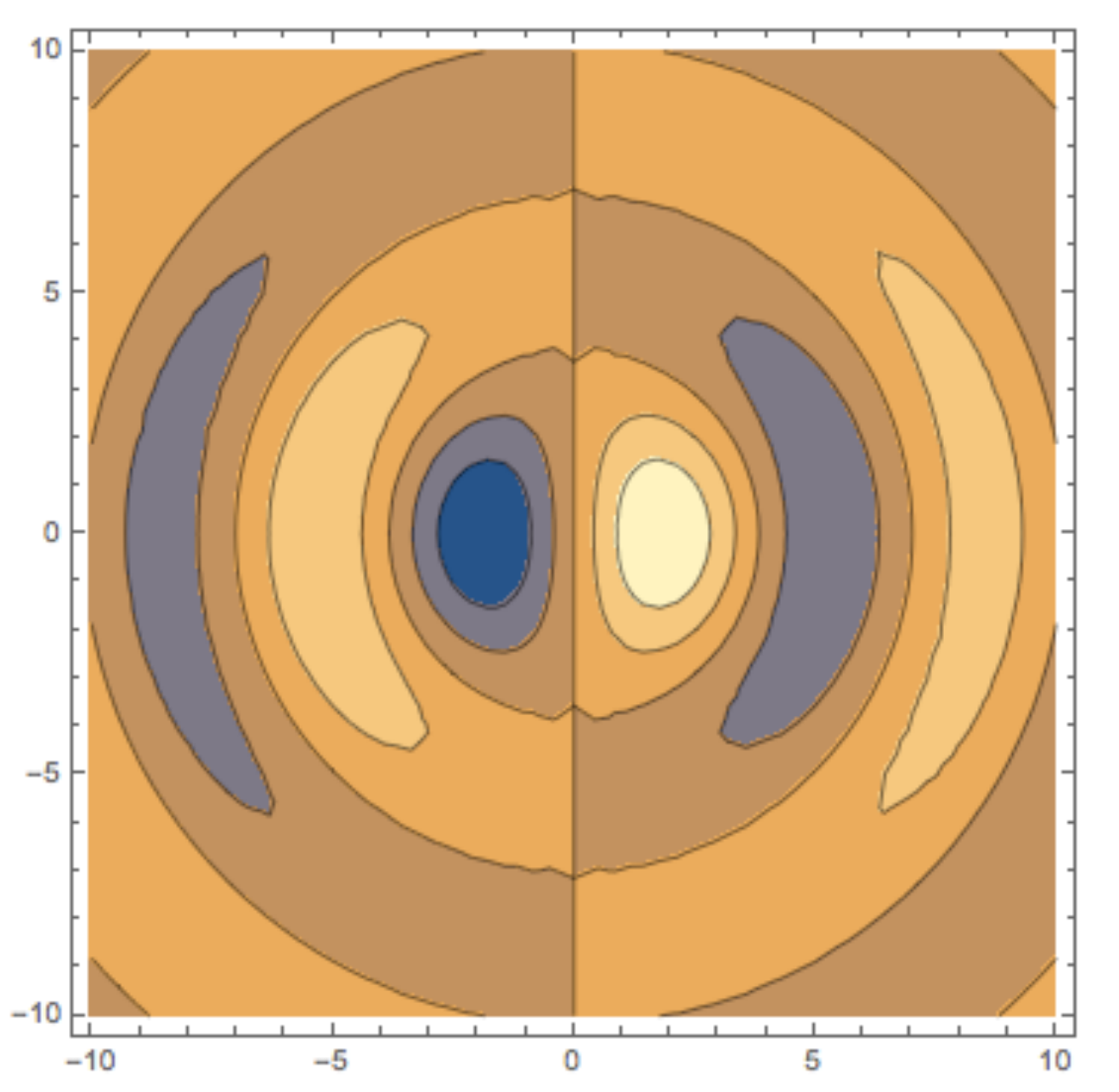}
		~~~~~\includegraphics[width=0.42\columnwidth]{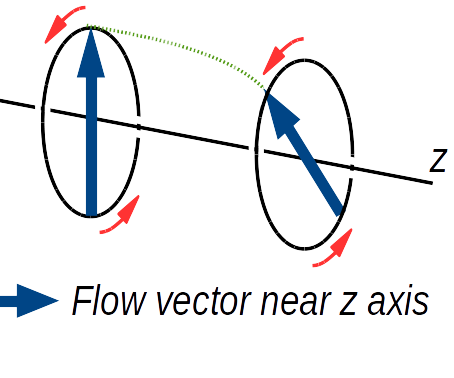}
	\caption{\em Cross section of the excess density of the phase vortex $\delta \rho_1 \propto J_1(r) \cos(\omega t-\theta-k_z z)$
	 at $t=z=0$ (left, courtesy Keith Moffatt). At the centre ($r=0 $), fluid is 
	 flowing towards $\theta=\pi/2$. The flow near the $z$ axis forms a helix (right).
	}
	\label{fig:j1}
\end{figure}

These experiments suggest a simple update of Maxwell's molecular vortex.
The density $\rho$ of Euler's compressible fluid obeys the wave equation to first order,
$\partial^2 \rho/\partial t^2 - c^2 \nabla^2 \rho=0$ where $c$ is the speed of sound,
and the update is a solution to the wave equation in cylindrical coordinates $(r, \theta, z)$
\begin{equation}
	\delta \rho_n ~~\propto~~ J_n (k_r r) \, \cos(\omega t - n \theta - k_z z)
	\label{eq:3d-phase-vortex}
\end{equation}
Here $\delta \rho_n$ is the excess density of the fluid, $n$ is an integer, 
$J_n$ is a cylindrical Bessel function of the first kind~\cite[\S9.1]{abramowitz1972handbook}, and 
$\omega^2 = c^2(k_r^2+k_z^2)$.
This is a phase vortex since the phase $S=\omega t - n \theta - k_z z$ 
advances by $-2 n \pi$ around the centre.
Figure \ref{fig:j1} shows the case $n=1$.

Mass flows around the axis on average because the 
flow speed \textbf{u} of the fluid is correlated with its density. 
From Euler's equation to first order,
$\rho_o \, \partial \textbf{u}/\partial t =-\nabla P=-c^2 \nabla \rho$
where we have used $c^2=dP/d\rho$, giving
$\rho_o \, \textbf{u}=-c^2 \int \nabla \rho \, dt$~\cite{faber1995fluid}.  
Substituting $\rho =\rho_o+R \cos (S)$ 
gives $\textbf{u}=\textbf{u}_o + c^2/(\omega \rho_o) \, R \cos (S) \, \nabla S$,
where  $\omega=\partial S/\partial t$ is constant;
here $\textbf{u}_o$ is not of interest in irrotational flow.
The remaining momentum per unit volume $\rho \, \textbf{u}$ has a term in $\cos^2(S)$, 
whose mean (neglecting any high-frequency variation) is $\frac12$, giving a mean momentum density~\cite{mcintyre1981wave}
\begin{equation}
	\pbar~~=~~\frac{c^2}{2 \omega \rho_o}  \, R^2 \, \nabla S
	\label{eq:define-pbar}
\end{equation}

By inspection of the phase $S=\omega t-n \theta-k_z z$, net mass flows around the phase vortex \eqref{eq:3d-phase-vortex}, $\oint \pbar.d\boldsymbol{\ell} \neq 0$. Maxwell's two equations
follow by defining $\textbf{B}=\nabla \times \pbar$ as shown above.

In general, if $f(\textbf{x},t)$ obeys the wave equation then so does $f(\textbf{x}', t')$ where the primed coordinates have undergone a Lorentz transformation with the characteristic speed $c$ of sound~\cite{dubois2011,barcelo2011analogue}. It follows that a moving line of force
is given by a Lorentz transformation of \eqref{eq:3d-phase-vortex} to first order. Note that a
moving magnetic field is likewise given by a Lorentz transformation of a stationary one, this time with $c$ being
the speed of light -- an apparent coincidence to which we will return.

Figure \ref{fig:j1} shows an elevated flow velocity near the centre.
This gives a negative Bernoulli pressure and a `tension' along the axis as in Maxwell's magnetic line of force.

Finally, the kinetic energy of (\ref{eq:3d-phase-vortex}) is unbounded
at large radius, from $J_1(r) \propto r^{-1/2} \, \cos(r+\pi/4)$ as $r \rightarrow \infty$.
In other fields of study, similar difficulties require renormalisation methods,
but they may be superfluous here.
Acoustic waves are usually adiabatic, with a reduced wave speed in 
regions of low pressure~\cite{faber1995fluid}.
This will reduce the speed near the axis and perturb the solution.
There is another system that obeys the same equations, 
namely 
an optical fibre. Refraction reduces 
the wave speed near the centre, and the waves decay exponentially with radius at large distance.

\section{Inverse square force}

Maxwell's 1861 paper did not have a good account of the electrostatic force~\cite{maxwell1861linesofforce}. 
Yet it was already known that a vibrating tuning fork 
attracts a balloon~\cite{forbes1881hydrodynamic}.
In 1880 Bjerknes investigated this by making two small bladders pulsate at the same frequency
in a tank of water. He observed an inverse-square force between them, a phenomenon now
known as the `secondary Bjerknes force'.
This seemed analogous to the electrostatic force and led
Lorentz to model electrons as `covibrating 
particles'~\cite{bjerknes1915hydrodynamische,sciam1882various,lorentz1902theory}. 

The force on a small body of volume $V$ is $V \nabla P$
where $P$ is the pressure.
Suppose one body, $A$, pulsates at angular frequency $\omega$, so pressure waves are transmitted through the fluid
and the pressure gradient near another body, $B$, varies as $\nabla P \propto \cos(\omega t)$.
If $B$ pulsates in phase, so its excess volume is also proportional to $\cos(\omega t)$,
then $V \nabla P$ has a term in $\cos^2(\omega t)$
which is never negative. The mean force is inverse-square
and the phase determines whether it is attractive or repulsive~\cite[p127]{faber1995fluid}. 

The same holds in a compressible fluid, where the pressure obeys the wave equation   
$\partial^2 P/\partial t^2 - c^2 \nabla^2 P=0$ to first order.
As noted above, this equation is symmetric under Lorentz transformation so that $\nabla P$, and hence the Bjerknes force, 
must have the same symmetry when the phases remain locally aligned. 
This lets us reuse the standard argument that extends the inverse-square force between stationary electrons to a moving frame.
Imposing Lorentz symmetry on an inverse-square force 
yields Maxwell's full equations of electromagnetism.

Oil droplets that are made to bounce
on a vibrating bath behave in this way~\cite{protiere2006particle}.
They bounce in the depressions, which tends to keep their phases aligned with the waves as above. 
Measurements show that they repel each other with an inverse-square force, while moving droplets experience an analogue of the magnetic force~\cite{bradyanderson2014droplets}.

\section{Electromagnetic waves}

We now revisit Maxwell's inspiration, namely Faraday's 1846 model of light 
as waves in lines of force (figure \ref{fig:faraday-line-force}).
In our update of his model, a disturbance or wavepacket travels along a phase vortex.
We will consider in detail the simplest 
case, amplitude modulation, where the Fourier components are given by \eqref{eq:3d-phase-vortex} with various
values of $k_z$. 
These components all have the same chirality, so the wavepacket has the same symmetry
as circularly polarised light.

The nonlinear terms in Euler's equation 
for a compressible fluid will perturb the shape of the disturbance
after it has propagated some distance.
Such processes are studied in fluid mechanics and typically result in
the disturbance being compressed; they are seen in phenomena
such as tidal bores and sonic booms~\cite{faber1995fluid}.

The wave described above is localised near the axis (see figure \ref{fig:j1}),
so it can only exchange its energy with a small system if 
the phase vortex passes close by. 
Einstein noted in 1905 that the energy density in light waves does not dilute with distance but it is absorbed discretely, consistent with our model~\cite{einstein1905heuristic}.

Maxwell envisaged a `sea' of lines of force, and noted they exert an inverse-square force
on average when they radiate from a central point~\cite[p160-163]{maxwell1855faraday}.
Regarding the average behaviour of the waves propagating along such lines, 
we saw from Bjerknes' result  that the motion obeys Maxwell's full equations of electromagnetism to first order.

This model extends to linearly polarised wavepackets, which inhabit lines of force that have equal and opposite chiral components, such as the following solution to the wave equation 
\begin{equation}
	\delta \rho_1 + \delta \rho_{-1} \propto 	J_1(k_r r) \cos(\omega t-k_z z) \cos(\theta-\theta_o)
\label{eq:linear-line-force}
\end{equation}
where the $\delta \rho_i$ are defined in \eqref{eq:3d-phase-vortex} and we have generalised
the origin of $\theta$ to be $\theta_o$. 
This line of force is oriented with an oscillating dipole of density parallel to $\theta=\theta_0$
along its length. 

We have described wavepackets in a fluid which have the same symmetry as polarised light.
Yet for many years polarised waves were thought impossible in a fluid. 
This belief arose from a misinterpretation of Fresnel's 1821 paper
on polarised light, as we now examine. 

Euler's fluid model of light fell from favour after Young and 
Fresnel could not think of a mechanism by which  
polarised waves might propagate in a fluid~\cite[p218]{cantor1981conceptions}\cite[p261]{buchwald1989rise}\cite{fresnel1821note}.
Some time after Fresnel's paper in 1821 it came to be assumed that
no such mechanism was possible.
Waves in vortices or phase vortices, such as those described above, do not appear to have been considered.  
Even a century later, Einstein maintained that some kind of `ether of the general theory of relativity' was needed, 
but could not see how it could be a fluid, given the belief 
that polarised waves are `not possible in a fluid'~\cite{einstein2007ether}.

\begin{figure}[htb]
	\centering
		\includegraphics[width=0.6\columnwidth]{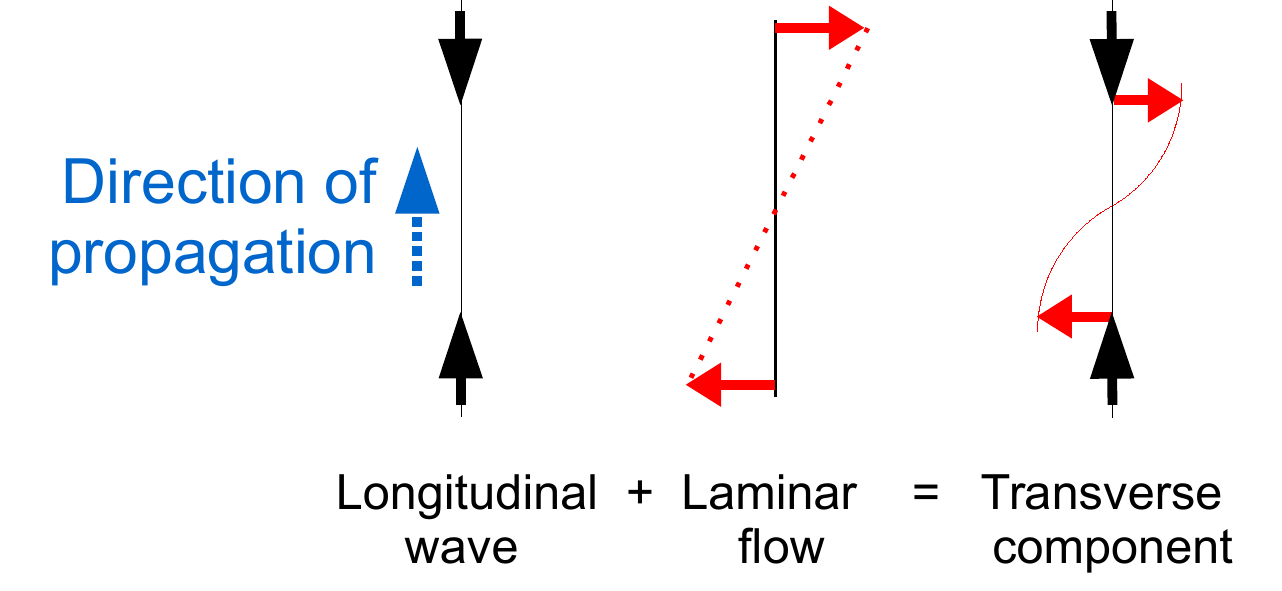}
	\caption{\em A wave displaces laminar flow}
	\label{fig:shear-flow}
\end{figure}

Yet today we have many examples of polarised waves in fluids. 
At its simplest, a wave propagating in laminar flow displaces the flow as shown in figure \ref{fig:shear-flow}. This makes it
asymmetric about the axis of a ray --
the historical meaning of `polarised'~\cite{cantor1981conceptions}.
Polarised waves are observed in the atmosphere, where the asymmetry is 
due to Coriolis forces~\cite{greenspan1990theory,moffatt1978field,Buhler2014b};
in superfluid $^4$helium where
the asymmetry is due to differential flows~\cite{adamenko2008transverse};
and superfluid
$^3$helium where it is due to atomic spin~\cite{landau1957oscillations,lee1999discovery,putterman}.
We saw more complex polarised waves arising from Maxwell's model. 

As for Maxwell, he did not know about the inverse-square secondary Bjerknes force 
and his 1861 fluid model did not have a good account of the electrostatic force.
He later downplayed it in favour of a more abstract field-theoretic presentation, similar to that taught today,
and fluid models of electromagnetism were not pursued. 

\section{The model of Clauser, Horne, Shimony and Holt}

A line of force such as \eqref{eq:linear-line-force} has a definite orientation, 
which determines the polarisation of any amplitude modulated wavepackets 
travelling along it.  
In this respect, our extension of Faraday's model of light
diverges from that advanced in 1969 by Clauser, Horne, Shimony and Holt (`CHSH'), 
in which they assumed that information relating to
polarisation can only be `carried by and localised within' a wavepacket or photon.
Basing a calculation on an earlier analysis by Bell, CHSH showed that their assumption leads to a contradiction to
the predictions of quantum mechanics which can be tested experimentally~\cite{bell1964einstein,clauser1969proposed}. 
When these `Bell tests' were conducted by
Freedman and Clauser in 1972 and then Aspect, Dalibard and Roger in 1982, they confirmed the quantum predictions~\cite{freedman1972experimental,aspect1982experimental}.

\begin{figure}[htb]
	\centering
		\includegraphics[width=0.3\textwidth]{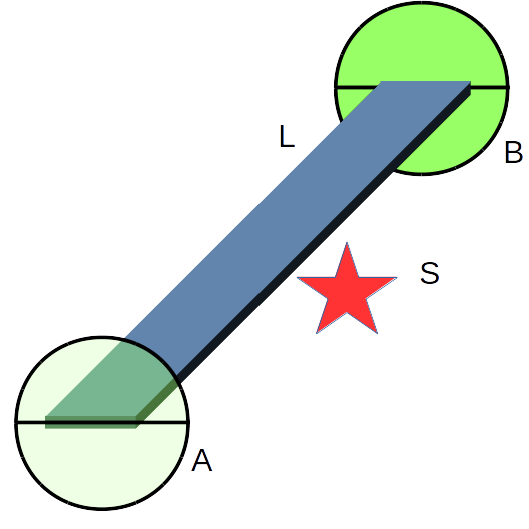}
	\caption{\em CHSH's experiment from the viewpoint of Faraday's approach. A source of light (S) stimulates a line of force (L) between two polarisers (A) and (B). }
	\label{fig:two-polarisers}
\end{figure}

Figure \ref{fig:two-polarisers} shows a CHSH experiment from the viewpoint of Faraday's approach.
In outline, a sudden disturbance $S$ stimulates a line of force $L$, causing wavepackets to travel 
along the line of force in
opposite directions until they reach  polarisers $A$ and $B$. Photomultipliers 
detect packets that pass through the polarisers. The figure shows the case where the polarisers and the line of force are oriented in the same direction, $\theta=0$ in the usual
cylindrical coordinates.

Suppose the polarisers contain particles (such as phase vortex rings~\cite{brady2013incompressible} 
or other quasiparticles~\cite{Volovik2003universe})
which are arranged with a net oscillating dipole of density in the
$\theta=\pi/2$ direction but no net dipoles along $\theta=0$. If they
couple to the lines of force in the vicinity then we would expect
the region between them to contain
two normal sets of lines of force, given by \eqref{eq:linear-line-force} with $\theta_o=\pi/2$ (which couple to or emanate from the dipoles) and $\theta_o=0$ (for example, lines emanating from further away 
to which the polarisers are transparent).
For simplicity, figure \ref{fig:two-polarisers} only shows a line of force with $\theta_o=0$.

In figure \ref{fig:two-polarisers} an oscillating system, or source $S$, stimulates a line of force oriented in the $\theta=0$ direction.
The stimulation occurs through the quadratic terms in Euler's equation 
(see~\cite{mead2000collective} and \cite{bradyanderson2013bell} for 
a parametric mechanism, which transfers the available energy completely or not at all). 
This creates amplitude modulated wavepackets
in the line of force, which travel in opposite
directions away from the source. 
They will pass unhindered
through the polarisers, which do not couple to them
because they have no dipoles parallel to $\theta=0$,
and will be detected by the photomultipliers
at both stations.
Conversely, if the source stimulates a line of force oriented in the $\theta=\pi/2$ direction,
then the wavepackets  will couple to the polariser and be absorbed or reflected, so neither photomultiplier
will register a signal.
Thus, the two signals will be 100\% correlated when the polarisers are parallel to each other.
This is observed experimentally.

Suppose the polariser $B$ in figure \ref{fig:two-polarisers} is now rotated through angle $\phi$
and the experiment is repeated. 
Near the source $S$, 
the lines of
force emanating from $A$ are oriented with $\theta_o=0$ or $\pi/2$, and those from $B$ with
$\theta_o=\phi$ or $\phi+\pi/2$.
Without
loss of generality we suppose the source couples parametrically to one of the lines of force from $A$ (the calculation is equivalent if it couples to
those from $B$).

First consider a wavepacket in a line of force which is oriented in the $\theta=0$ direction. As before, 
one of the wavepackets will travel towards $A$ and will
pass directly through the polariser and 
trigger the detector.
The other wavepacket travels in the opposite direction, and will propagate for some
distance near the lines
of force from $B$.
The wavepacket has a component in the $\phi$ direction whose amplitude is proportional to $\cos \phi$, and in the $\phi+~\pi/2$ direction it is proportional to $\sin \phi$.
These components will couple to the lines of force from $B$ through nonlinearities in the equation of motion. 
There are quadratic terms in Euler's equation, which, to lowest order, give a probability
of resonant transfer which is quadratic in amplitude.
See~\cite{bradyanderson2013bell} for an example parametric mechanism for such coupling.
It follows that the lines of force at angle $\phi$ 
will be stimulated with probability $\cos^2 \phi$, and those at angle 
$\phi+\pi/2$ will be stimulated with
probability $\sin^2 \phi$.

Thus, with suitable normalisation, the number of times a wavepacket is detected
at both $A$ and $B$ is $N_{++}=\cos^2 \phi$ and the number of times it is detected at $A$ but not $B$ is $N_{+-}=\sin^2 \phi$. From symmetry we also have $N_{-+}=N_{+-}$ and
$N_{--}=N_{++}$ in the obvious notation.
We are interested in the correlation function

\[
	\begin{aligned} \frac{N_{++} + N_{--}  -N_{+-}  -N_{-+} }
	{N_{++}+ N_{--} +N_{+-}+N_{-+} }
	& = \frac{\cos^2 \phi - \sin^2 \phi}{\cos^2 \phi + \sin^2 \phi} \\
	& = \cos( 2 \phi) \end{aligned}
\]
which is the same as the quantum prediction (Bell's calculation of $\cos \phi$ was for spin-half particles~\cite{bell1964einstein},
rather than spin 1 photons as here).
The later `Bell test' experiments confirmed the quantum prediction~\cite{freedman1972experimental,aspect1982experimental}.
They have been extended in a number of ways, but they have only been conducted using stationary polarisers~\cite{vervoort2013bell,vervoort2014no}.

These experiments challenge the CHSH assumption that information about polarisation can only be
`carried by and localised within' the photons.
The conventional approach holds that the CHSH assumption is true but incomplete:
further hypotheses are made about parallel
universes or about non-local phenomena that can transmit information
(but not actual observable signals) faster than light~\cite{bell2004speakable,everett2012everett}.
Our extension of Faraday's 1846 model of light offers a simpler alternative that is both physical and local.
The CHSH assumption is not true in Faraday's model.
Instead there is 
prior communication of orientation
along phase vortices such as \eqref{eq:linear-line-force}, 
communication which the CHSH calculation excludes by its explicit assumption.

\section{Conclusion}

In 1746 Euler modelled light as waves in a compressible fluid.
Most nineteenth-century scientists rejected his model because they believed 
polarised waves to be impossible in a fluid, a belief that 
is now well known to be false.

We brought Maxwell's 1861 model of a magnetic line of force up to date using modern knowledge of polarised waves and of experiments on quantised magnetic flux.
Our model obeys the equations for Euler's fluid and supports light-like solutions which are polarised, 
absorbed discretely, consistent with the Bell tests,
and obey Maxwell's equations to first order.

Euler's fluid obeys the wave equation to first order. We saw that this equation is symmetric
under Lorentz transformation, so if the fundamental particles
are quasiparticles in such a fluid 
then the Lorentz symmetry of material bodies
emerges naturally. Euler's fluid also has the symmetries 
of general relativity~\cite{barcelo2011analogue}, which has led to 
experiments on Hawking radiation in a superfluid~\cite{unruh1981experimental,lahav2010realization}.
For quasiparticles, see phase vortex rings~\cite{brady2013incompressible} and Volovik's 
model~\cite{Volovik2003universe}.

Finally, for further connections with quantum mechanics, see~\cite{bradyanderson2014droplets,couder2006single} 
for experiments in which bouncing droplets exhibit quantum-like phenomena.
The main reason such fluid analogues are not considered more widely 
is the assumption that quantum mechanics simply cannot emerge from classical
phenomena -- principally because nobody had been able to think of a classical model of light that is consistent with Maxwell's equations and
reproduces the Bell test results quantitatively. Our extension of Faraday's model of light
provides a counterexample.

In general, classical models of quantum phenomena must feature long-range order
if they are to be consistent with the Bell tests; see  Vervoort~\cite{vervoort2013bell,vervoort2014no}.
With bouncing droplets in two dimensions, this order arises from the 
driving oscillation~\cite{protiere2006particle,bradyanderson2014droplets}. 
In the three-dimensional model shown here, the order comes from a line of force. 
Some variants of 't Hooft's cellular automaton interpretation of quantum mechanics may likewise have such order~\cite{thooft2014fa}, or it may emerge from the synchronisation of the spins of particles~\cite{bradyanderson2013bell}. No doubt there are other possibilities. 
But given the mechanism we have described, it is indeed possible for quantum mechanics to emerge from an underlying classical system.

\subsection*{Acknowledgement}

We thank Robin Ball, Michael Berry, Basil Hiley, Michael McIntyre, Keith Moffatt, 
Theo Nieuwenhuizen, Simon Schaffer, Bill Unruh, Louis Vervoort, Grisha Volovik and attendees at the fluids
seminar at Cambridge for helpful discussions.

\ifnature
\else
	{\footnotesize \bibliography{book}
	\bibliographystyle{unsrt}}
\fi

\end{document}